\documentclass{emulateapj}
\usepackage{hyperref}
\hypersetup{
colorlinks=true,
linkcolor=red,
citecolor=blue,
filecolor=cyan,
urlcolor=magenta,
}

\usepackage{graphicx}
\usepackage{natbib}
\usepackage{multirow}
\usepackage{subfigure}

\submitted{$\qquad$}

\begin{document}

\title[Search for global $f$-modes and $p$-modes in the $^8$B neutrino flux]{
Search for  global $f$-modes and $p$-modes in the $^8$B neutrino flux}
\author{Il\'\i dio Lopes\altaffilmark{1,2}}

\altaffiltext{1}{Centro Multidisciplinar de Astrof\'{\i}sica, Instituto Superior T\'ecnico, 
Universidade de Lisboa, Av. Rovisco Pais, 1049-001 Lisboa, Portugal;ilidio.lopes@ist.utl.pt,} 
\altaffiltext{2}{Departamento de F\'\i sica, Escola de Ci\^encias e Tecnologia, 
Universidade de \'Evora, Col\'egio Luis Ant\'onio Verney, 7002-554 \'Evora - Portugal;ilopes@uevora.pt} 

% % % % % % % % % % % % % % % % % % % % % % % % % % % % % % % % % % % % % % %
\date{Received 2013 May 16; accepted 2013 September 15}

\begin{abstract} 
The impact of global acoustic modes  on the $^8$B neutrino flux time 
series is computed for the first time. 
It is shown that  the time fluctuations of the $^8$B neutrino flux depend on 
the amplitude of acoustic eigenfunctions  in the region where the $^8$B neutrino flux is produced: 
modes with low $n$ (or order) that have eigenfunctions 
with a relatively large amplitude in the Sun's core, strongly affect the  neutrino flux; 
conversely,  modes with high $n$  that have eigenfunctions with a minimal amplitude in the Sun's 
core have a very small impact on the neutrino flux.  
It was found that the global modes with a larger impact on the $^8$B neutrino flux 
have a frequency of oscillation in the interval $250\;\mu{\rm Hz}$ to $500\;\mu{\rm Hz}$
(or a period in the interval $30$ minutes to $70$ minutes), such as 
the  $f$-modes  ($n=0$) for the low degrees, radial modes of order $n\le 3$, 
and the dipole mode of order $n=1$.
Their corresponding neutrino  eigenfunctions are very sensitive to the  solar inner core 
and are unaffected by the variability of  the external layers of the solar surface. 
If time variability of neutrinos is observed for these modes, it will lead to new ways of improving the sound 
speed profile inversion in the central region of the Sun.
\end{abstract}

%\pacs{ln}
\keywords{elementary particles -– neutrinos -– nuclear reactions, nucleosynthesis,
abundances -– Sun: general -– Sun: helioseismology -– Sun: oscillations}

\maketitle

% 
%%%%%%%%%%%%%%%%%%%%%%%%%%%%%%%%%%%%%%%%%%%%%%%%%%%%%%%%%%%%%%%%%%%%%%%%%%%%%
%
%%%%%%%%%%%%%%%%%%%%%%%%%%%%%%%%%%%%%%%%%%%%%%%%%%%%%%%%%%%%%%%%%%%%%%%%%%%%%
\section{Introduction}
%%%%%%%%%%%%%%%%%%%%%%%%%%%%%%%%%%%%%%%%%%%%%%%%%%%%%%%%%%%%%%%%%%%%%%%%%%%%%

Over the last three decades, the neutrino physics community 
has been able to characterize some of 
the basic properties of neutrinos, namely,  
their oscillatory nature between flavors in a vacuum as well 
as in matter~\citep[][]{1958JETP....6..429P,1978PhRvD..17.2369W,1986NCimC...9...17M}, 
and fine-tune the properties of  the nuclear reactions 
where neutrinos are produced in the Sun's core. 
Although some important questions remain unsolved,
the basic theory of neutrino physics is by now well understood~\citep[e.g.,][]{2010LNP...817.....B}.

The high quality of neutrino data available from current and future
experiments~\citep[e.g,][]{2011PhRvD..83e2010A,2010PhRvD..82c3006B} 
will provide a way to make accurate diagnostics of the solar nuclear region,  
allowing  us to test the shortcomings and limitations of 
the present standard solar model~\citep[SSM;][]{1993ApJ...408..347T,2010ApJ...715.1539T,2013ApJ...765...14L}.

Our understanding of the physical processes occurring in the Sun's interior has
progressed significantly due to the use of two complementary solar probes: solar neutrino 
fluxes and helioseismology data. In particular,    
it was possible to infer some of the basic dynamic properties 
of plasma in the Sun's core, like the ones caused by the presence  of magnetic fields, 
rotation and other flow motions~\citep[e.g.,][]{2011RPPh...74h6901T}.
Furthermore, this combined analysis of data was also used to
test fundamental concepts in physics~\citep[e.g.,][]{2003MNRAS.341..721L,2012ApJ...745...15C}
or to study the properties of dark
matter~\citep[e.g.,][]{2002PhRvL..88o1303L,2010Sci...330..462L,2010ApJ...722L..95L,2012ApJ...752..129L}. 
In the latter case, the Sun is used as a cosmological tool, 
following a research strategy quite common in modern cosmology of using stars 
to constrain the properties of dark matter~\citep[e.g.,][]{2011PhRvL.107i1301K,2013ApJ...765L..21C}. 

Helioseismology, by means of sophisticated inversion techniques,  
has been able to infer some fundamental quantities 
related to the thermodynamics and dynamics properties of the Sun's interior~\citep{2011RPPh...74h6901T}.
The two leading results worth mentioning are  speed of sound~\citep{2009ApJ...699.1403B,2001ApJ...555L..69T,2004PhRvL..93u1102T} 
and the differential rotation~\citep{1996Sci...272.1300T} profiles. These results are obtained  between the layers $0.8 R_\odot $ (beneath the photosphere)  and  $0.1 R_\odot $ (nuclear region). 
This achievement was possible due to  high precision frequency measurements of several thousands 
of vibration acoustic modes with a precision better than 
$10^{-5}$~\citep[see][and references therein]{2011RPPh...74h6901T}. 
Unfortunately, the physics in the solar inner core is still largely unknown  
due to the fact that only a small number of acoustic modes 
are sensitive to that region and among these only a few have been successfully
detected~\citep{2000ApJ...537L.143B,2001SoPh..200..361G,2009ApJS..184..288J}.  
Naturally, the diagnostic of the solar core could improve significantly 
if the discovery of gravity modes is
confirmed~\citep{2004ApJ...604..455T,2007Sci...316.1591G,2012ApJ...746L..12T}.

Solar neutrino flux time fluctuations have been studied for several decades,
mainly focusing on finding evidence of the solar magnetic cycle and day--night earth
effects~\citep[e.g.,][]{2004APh....21..511L,2005PhRvD..72k3004S,2005PhLB..623...80F,2006JHEP...02..035C}.
 
This work focuses on the impact of  global acoustic mode oscillations
on solar neutrino fluxes. 
Some of these acoustic modes propagate to the deepest layers of 
the Sun's nuclear region, and perturb the local temperature.
Global acoustic modes with small order 
are very sensitive to the central core and are not perturbed by the surface. 
The space instrument Global Oscillations at Low Frequencies (GOLF) has begun to detect some of these modes although not all 
of them have yet been identified due their small amplitudes at the surface. 
See Table~\ref{tab:1} and the review~\citep{2012RAA....12.1107T} which gives a synthesis 
of the performances of these modes.

In principle, time variations of neutrino flux 
measurements could provide a way to measure the frequencies 
of such modes with a very small amplitude at the surface, 
without the variability related to the complex structure of the solar upper layers,
caused by non-adiabaticity, turbulent convection, sub-photospheric differential rotation, 
magnetic activity and solar oblateness~\citep{2009LRSP....6....1H}.

In this Letter, we discuss the possibility of detecting global acoustic oscillations
of low order in solar neutrino flux fluctuations.
It is  true  that global modes of very high order have a  small amplitude  in the  Sun's core, 
but global modes of low order have quite a large amplitude in the core as well as on the surface.    
There is the distinct possibility of observing them in the solar neutrino flux fluctuations,  
or in the case that the observations are not confirmed, 
it will be possible to put an upper bound on their amplitude in the Sun's core. 
We estimate for the first time the impact of  global acoustic 
modes in the  $^8$B neutrino flux and predict the  global modes which are more likely  
to be found in solar neutrino flux time series.

% - main results with solar neutrino physics.
%- Acoustic modes
%%%%%%%%%%%%%%%%%%%%%%%%%%%%%%%%%%%%%%%%%%%%%%%%%%%%%%%%%%%%%%

\begin{figure}
\centering
\includegraphics[scale=0.5]{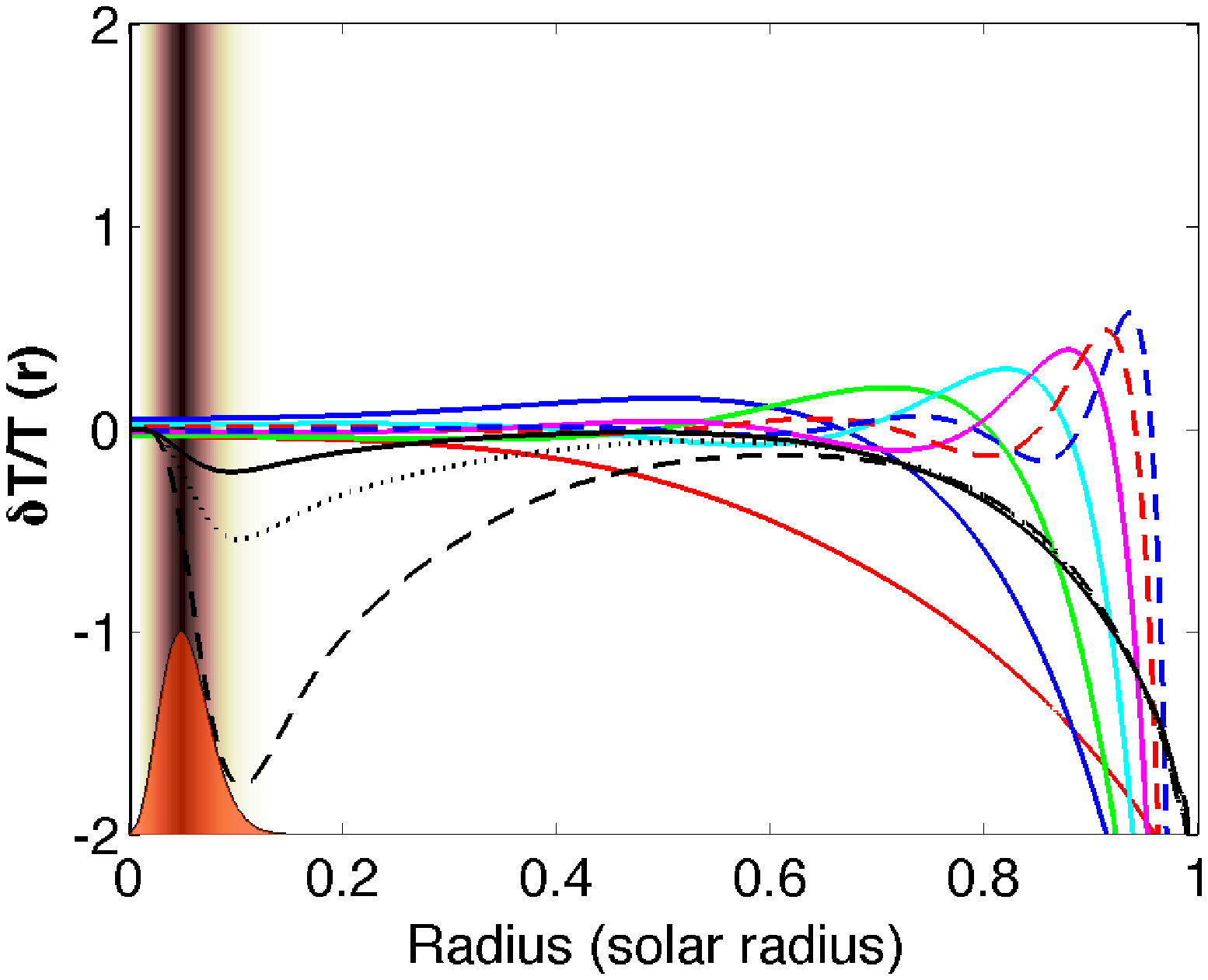}
\includegraphics[scale=0.5]{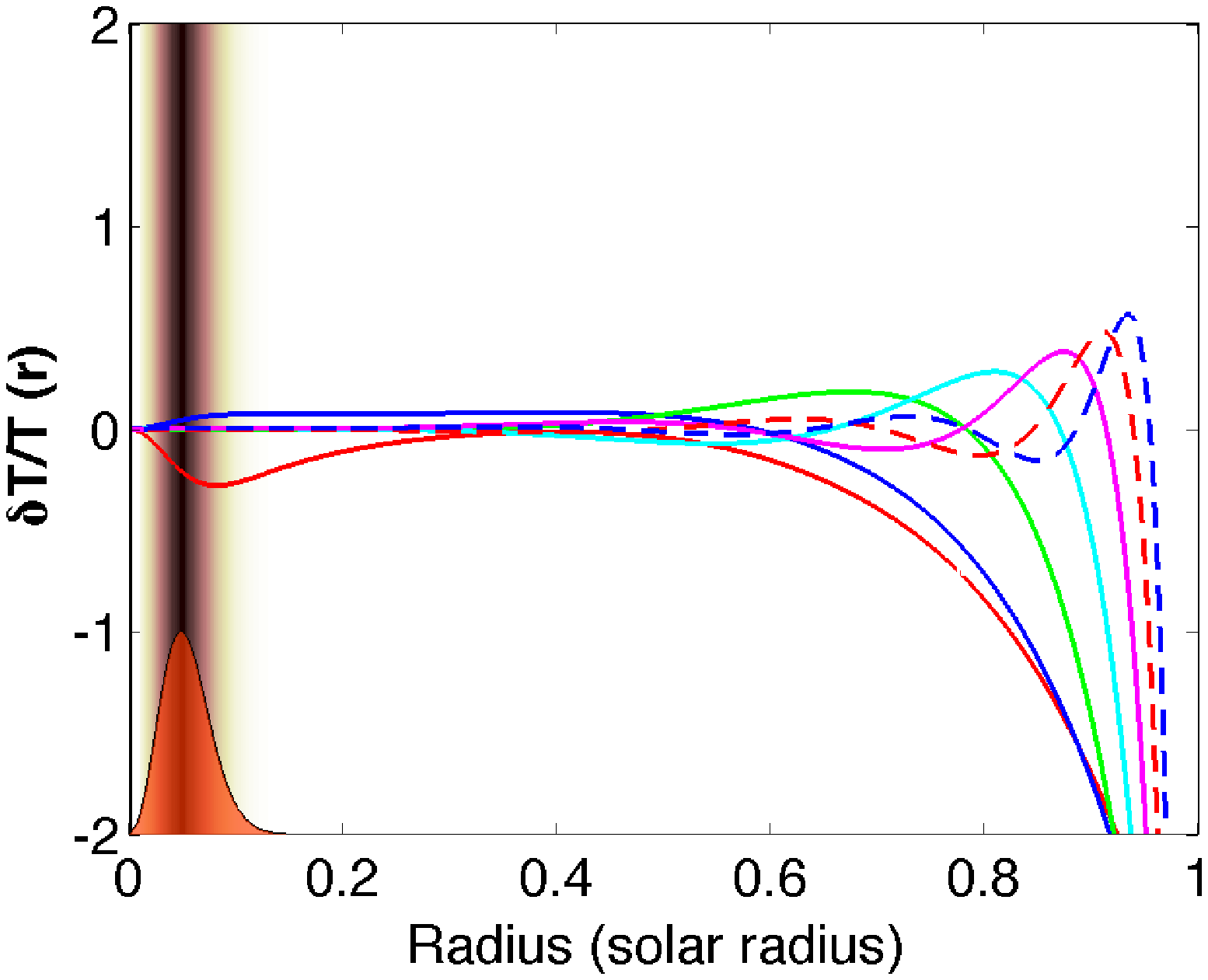}
\caption{The region of emission of $^8$B neutrinos (brown dashed area) and their flux $\Phi$ (r) (orange area). The $(\delta T/T)_{ln}$ eigenfunctions are drawn
for selected  radial ($l=0$) and quadrupole ($l=2$) acoustic modes computed for the current solar
model: 
{\bf (a)} $p_1,\cdots,p_7$ radial ($l=0$) modes + $f$ modes for $l=3,4,5$;
{\bf (b)} 
$f$ and
 $p_1,\cdots,p_6$ quadrupole ($l=2$) modes. The color for modes $l=0$ and $l=2$ runs from the lower to 
the higher value of $n$ in the following sequence:  red,  blue, green, cyan, magenta, dashed red and dashed blue %The line style for 
The $f$ modes of degree $l=3$, $l=4$ and $l=5$ correspond to continuous, 
dotted and dashed  back curves, respectively. 
For convenience, $(\delta T/T)_{ln}$ of the $l=0$ and $l=2$ modes were normalized to the value 4,
the $f$ modes of $l=3$, $l=4$ and $l=5$ to the value 8  and $\Phi (r)$  to the value  1.}
\label{figu1}
\end{figure}

\begin{table} %[htb]
\centering
\caption{$^8$B Neutrino Flux Variations \\ Caused by Acoustic Modes}
\begin{tabular}{rlllc}
\hline
\hline
Mode~\footnote{The observational frequency table is obtained from 
a compilation made by~\citet{2012RAA....12.1107T}, after the observations of ~\citet{2000ApJ...537L.143B,2001SoPh..200..361G,2004ApJ...604..455T,2009ApJS..184..288J}.}
&  Frequency [obs]    & Frequency  [th]   &   ${\cal B}_{ln}$    \\
$(l,n)$\footnote{Mode classification scheme (for a fixed $l$): 
$n=0$ is the $f$-mode and  $n\ge 1$ are the $p_n$-modes~\citep[e.g.,][]{1989nos..book.....U}.}
       &  ($\mu Hz$)    &   ($\mu Hz$)     &   $\qquad$ \\
\hline
$(l=\cdots,\; n=0)$        &                         &        & $ \times  10^{0}$   &    \\
$2$         $f$     &  $-$                            & $347.10 $  &  $0.8069$  \\
$3$         $f$     &  $-$                            & $386.65$   &  $0.9704$  \\
$4$         $f$     &  $-$                            & $405.43$   &  $2.0608$  \\
$5$         $f$     &  $-$                            & $415.08$   &  $5.7283$  \\
\hline
 % Table 1 - Turck-Chieze & Lopes
 $(l=0,n=\cdots)$     &       &      & $ \times 10^{-1}$    &    \\
 $p_1$    &  $258.60\pm 0.030$     & $258.43 $   &  $1.5280$   \\
 $p_2$  &  $-$                             & $404.36$       &  $1.9924$     \\
$p_3$  &  $535.75\pm 0.010 $    & $535.57 $     &  $1.4123$    \\
$p_4$   &  $-$                             & $679.87 $   &  $0.9378$      \\
$p_5$   &  $825.23\pm 0.030$     & $824.51 $   &  $0.6264$   \\
$p_6$   &  $972.612\pm 0.005$   & $971.76 $  &  $ 0.4376$     \\
$p_7$   &  $-$                             & $1116.94 $  &  $0.3221$  \\
 $p_8$   & $1263.215\pm 0.01$    & $1262.31 $   &  $0.2336$    \\
 $p_9$   & $1407.49\pm 0.01$      & $1406.39 $    &  $0.1721$    \\
 $p_{10}$  & $1548.304\pm 0.01$    & $1547.44 $ &  $0.1238$        \\
\hline
 $(l=1,n=\cdots)$      &       &      & $ \times 10^{-2}$   &    \\
 $p_1$    &  $-$                              & $283.80 $  &  $21.674$   \\
 $p_2$   &  $-$                               & $448.38$   &  $2.5458$     \\
 $p_3$   &  $- $                              & $596.67 $ &  $1.6681$    \\
 $p_4$   &  $-$                              & $746.41 $ &  $1.2171$      \\
 $p_5$   &  $-$                              & $893.19 $  &  $0.9868$     \\
 $p_6$   &  $1039.465\pm 0.003$   & $1038.95$  &  $0.7835$  \\
 $p_7$   &  $1185.60\pm 0.05$    & $1184.84 $   &  $0.6448$    \\
 $p_8$   & $1329.63\pm 0.01$   & $1328.87 $    &  $ 0.5188$     \\
$p_9$    & $1472.86\pm 0.02$    & $1472.20 $  &  $0.4065$      \\
$p_{10}$   & $1612.746\pm 0.01$    & $1611.99 $  &  $0.3210$      \\
\hline
 $(l=2,n=\cdots)$        &       &      &   $ \times  10^{-3}$  &    \\
 $p_1$     &  $-$                               & $382.26 $    &  $177.67$   \\
 $p_2$    &  $-$                               & $514.48$      &  $5.2266 $     \\
 $p_3$    &  $- $                              & $664.06 $   &  $0.6738$   \\
 $p_4$    &  $-$                               & $811.33 $   &  $0.2446 $     \\
 $p_5$    &  $-$                               & $959.23 $ &  $0.5623$     \\
 $p_6$    &  $-$                              & $1104.28$    &  $0.6874$   \\
 $p_7$    &  $-$                              & $1249.78 $   &  $0.7097 $   \\
 $p_8$    & $1394.68\pm 0.01$        & $1393.68 $   &  $0.6865$      \\
 $p_9$     & $1535.865\pm 0.006$      & $1535.08 $  &  $0.6138$   \\
 $p_{10}$   & $1674.534\pm 0.013$    & $1673.80 $   &  $0.5455$     \\
\hline
 $(l=3,n=\cdots)$       &       &      & $ \times 10^{-4}$   &    \\
$p_1$      &  $-$                            & $415.95$   &  $517.5$   \\
$p_2$    &  $-$                            & $564.69$   &  $15.608 $    \\
$p_3$     &  $- $                           & $718.30 $  &  $ 4.2462 $     \\
$p_4$     &  $-$                            & $866.47 $ &  $1.6805$     \\
$p_5$       &  $-$                            & $1014.41 $  &  $0.5456 $     \\
$p_6$       &  $-$                            & $1160.82 $  &  $0.0577 $    \\
$p_7$     &  $-$                            & $1305.85 $    &  $0.4158$   \\
$p_8$       & $-$                           & $1450.10 $   &  $0.6137 $      \\
$p_9$       & $-$                          & $1590.63 $ &  $0.7015$       \\
$p_{10}$     & $1729.74\pm 0.02 $   & $1728.34 $  &  $0.7355 $    \\
\hline
\hline
\end{tabular}
\label{tab:1}
\end{table}

% Ilidio 
%%%%%%%%%%%%%%%%%%%%%%%%%%%%%%%%%%%%%%%%%%%%%%%%%%%%%%%%%%%%%%%%%%%%%%%%%%%%%
\section{Propagation of acoustic waves towards the core of the Sun}
%%%%%%%%%%%%%%%%%%%%%%%%%%%%%%%%%%%%%%%%%%%%%%%%%%%%%%%%%%%%%%%%%%%%%%%%%%%%%

As we are   looking for the propagation of acoustic waves generated in the
upper layers of the convection zone by 
turbulence~\citep[e.g.,][]{1994ApJ...424..466G}, 
any acoustic oscillation is regarded as a perturbation of 
the internal structure of the star.
Therefore, any perturbed thermodynamic quantity $f$ is such that $f=f_o+\delta f$, 
where $f_o$ and $\delta f $ denote the equilibrium value and the 
Lagrangian
perturbation of $f$.~\footnote{Following standard perturbation analysis, 
$\delta f $ in  spherical coordinates,
reads  $\delta f \propto e^{-i\omega t - \eta t} $   
where $\omega $  and $\eta$ are the frequency 
and the damping rate~\citep[e.g.,][]{1989nos..book.....U}.}
It follows that for each eigenmode of vibration of degree $l$ and order 
$n$ (where the contribution of the magnetic field and rotation is considered negligible) 
that there is an unique  eigenfrequency $\omega_{ln}$ and an unique set of eigenfunctions~\citep[e.g.,][]{1989nos..book.....U}. 
For example,  the perturbation of temperature will have an unique radial 
eigenfunction $\delta T_{ln}(r)$.\footnote{In this case, 
$n$ is a positive integer that relates with
the number of nodes of $\delta T_{ln}(r)$.
Following the usual classification scheme for modes with the same degree $l$: 
$n=0$ is the $f$-mode and  $n\ge 1$ are the $p_n$-modes~\citep[e.g.,][]{1989nos..book.....U}.} 
Table~\ref{tab:1} shows the frequencies of the radial and non-radial 
acoustic modes computed for the current  SSM~\citep{1993ApJ...408..347T,1997A&AS..124..597M,2010ApJ...715.1539T,2013ApJ...765...14L}.
This solar model is identical to others published in the literature~\citep[e.g.,][]{2011ApJ...743...24S}.
As shown in Table~\ref{tab:1}, the difference between theoretical $[th]$ and observational $[obs]$ frequencies
is smaller than $0.1\%$  \citep{2000ApJ...537L.143B,2001SoPh..200..361G,2004ApJ...604..455T,2009ApJS..184..288J}.

Figure~\ref{figu1} shows $\delta T_{ln}(r)$  for $p_n$ modes with $l=0$ and  $l=2$
and the $f$ modes for $l=2,\cdots,5$. Global acoustic modes based on the eigenfunction behavior can be classed into two types: modes of high $n$ -- for which the amplitude  of the radial eigenfunction 
is very small near the center of the star and its maximum occurs near the surface; 
and modes of lower $n$  --  for which the amplitude of the radial eigenfunction is quite large
near the center of the star. Examples of the latter type are $p_1$ and $p_2$
radial, dipole and quadrupole modes, as well as the $f$ quadrupole and octapole modes.
Somehow the eigenfunctions of such modes of low $n$ in the Sun's core exhibit a behavior 
identical to that of the eigenfunctions of gravity modes~\citep{2000A&A...353..775P}.

\begin{figure}
\centering
\includegraphics[scale=0.45]{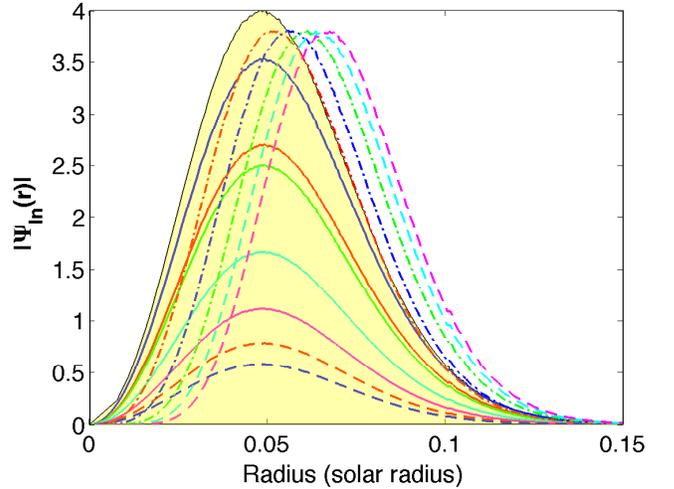}
\caption{ 
The   modulus of  {\it neutrino survival  eigenfunctions}  $|\Psi_{ln}|$,
for a selected  number of acoustic modes (see Table~\ref{tab:1}):
{(\bf a)} $p_1,\cdots,p_7$ radial ($l=0$) modes corresponding to
the same temperature eigenfunctions shown in Figure~\ref{figu1}(a) (the same color scheme).
{(\bf b)} $p_1$ dipole ($l=1$) mode ($-.-$ red line),  
$f$ quadrupole ($l=2$) mode ($-.-$ blue line),
$f$ octupole ($l=3$) mode ($-.-$ green line)
$f$  mode for $l=4$ ($- -$ cyan line)
$f$  mode for $l=5$  ($- -$ magenta line).
$|\Psi_{ln}|$ is a  dimensionless quantity computed using Equation (\ref{eq:Psi_ln}).
The source of $^8$B neutrino flux, $\Phi (r)$, is shown  as a light-yellow area.
The functions  $\Phi(r)$ are normalized to $4$ to facilitate the comparison 
with other curves. Similarly, the functions $|\Psi_{ln}|$ in the case {(\bf b)}
are normalized to 3.8.} 
\label{figu2}
\end{figure}

\begin{figure}
\centering
\includegraphics[scale=0.50]{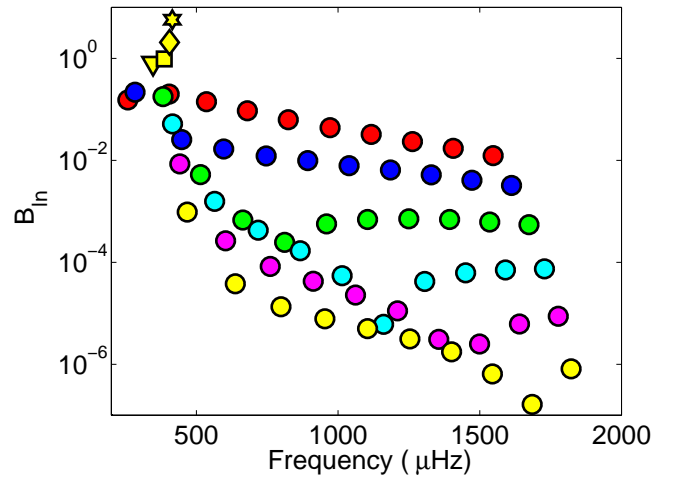}
\caption{${\cal B}_{ln} $  for global modes:
{\bf (a)} $p_n$ ($n=1,\cdots,10$) for $l=0$ (red bullet), $l=1$   (blue bullet), 
$l=2$  (green bullet),  $l=3$ (cyan bullet), $l=4$ (magenta bullet) and $l=5$  (yellow bullet). 
{\bf (b)} $f$-mode for $l=2$ (yellow triangle),  $l=3$ (yellow square),  $l=4$ (yellow diamond)
and $l=5$ (yellow hexagram).    $B_{ln} $ is computed as proposed by Equation (\ref{eq:B_ln}). 
See Table~\ref{tab:1} for numerical values of ${\cal B}_{ln} $. }
\label{figu3}
\end{figure}

The propagation of acoustic waves toward  the Sun's center  perturbs the local
thermodynamic structure, such as the chemical abundances, the 
density $\rho$ and the temperature $T$, triggering fluctuations in the energy generation rate $\epsilon$
of the different nuclear reactions, in particular the ones
related to the production of solar neutrinos. Therefore, 
the neutrino flux $\Phi(r)$ reads~\citep[e.g.,][]{1994sse..book.....K}:
\begin{equation}
\frac{\delta \Phi}{\Phi} = 
\frac{\delta \rho}{\rho} +\vartheta \frac{\delta T}{T} \simeq \zeta_o \frac{\delta T}{T},
\label{eq:a}
\end{equation}   
where $\vartheta $ and $\zeta_o$ are constants.
In the derivation of this result, $\delta \Phi$\footnote{In the remainder of this Lettter, 
$\delta \Phi (r)$ reads $\delta \Phi$, unless stated otherwise. 
Similarly $\Phi_o$  (as computed for the SSM) reads $\Phi$. 
The same rule applies to other thermodynamical quantities.} is considered 
proportional to $\delta \epsilon$, such that $\delta \Phi/\Phi\approx \delta \epsilon/\epsilon$.
The second (approximated) equality of Equation (\ref{eq:a}) is obtained as follows:
for adiabatic non-radial acoustic oscillations (which are valid in the interior of the Sun)
the following relation is valid  
${\delta\rho}/{\rho} =\left(\Gamma_3-1\right)^{-1} {\delta T}/{T}$,
where $\Gamma_3-1=\left(\partial T /\partial \rho  \right)_s $ is 
the derivative being taken at constant specific entropy $s$~\citep[e.g.,][]{1989nos..book.....U}. 
In the Sun's core, as the plasma 
is fully ionized, $\Gamma_3=5/3$.  
It is also assumed that the average cross-section between reagent particles
  is proportional to the $T^{\vartheta}$ 
where $\vartheta$ is an exponent,
and the perturbation of the mass fractions of the reacting particles 
are considered negligible. In particular, the production of the $^8$B neutrino flux, 
$\Phi$, is estimated to be proportional to $T_c^{24}$  where $T_c$  
is the temperature at the center of the Sun~\citep{1996PhRvD..53.4202B}.
Accordingly, if the value of $\vartheta$ is taken to be equal to $24$
then $\zeta_o\equiv 3/2+\vartheta=25.5$. 
This result shows that the relative variation on the neutrino flux is $25.5$ 
times larger than the temperature perturbation that has caused it. 

It follows that the total neutrino flux when perturbed by an
acoustic mode of frequency $\omega_{ln}$ is given by 
\begin{equation}
\phi_{ln}(t)=\phi_o+ \Delta \phi_{ln} (t),
\label{eq:phi_neutrino} 
\end{equation} 
where $ \phi_o$ is the total neutrino flux (independent of time) as computed by 
the SSM (equilibrium model)
and $\Delta \phi_{ln}(t)$ is the amount of neutrino flux produced or suppressed 
by the acoustic mode of vibration.  By integrating the 
Equation  (\ref{eq:a}) for the total mass of the star,  $\Delta \phi_{ln} (t)$ reads
\begin{equation} 
\Delta \phi_{ln}(t)  ={\cal A}_{ln}\;\zeta_o \;{\cal B}_{ln}\;\phi_o \;\;e^{-i\omega_{ln} t} \; e^{-\eta_{ln} t}, 
\label{eq:delta_phi_t}
\end{equation}  
where  ${\cal A}_{ln}$ is an amplitude related to the excitation source of global modes 
and ${\cal B}_{ln} $ is a quantity  that determines the fraction of the total neutrino flux which
is affected by  the mode. ${\cal B}_{ln} $ reads
\begin{equation} 
{\cal B}_{ln}=\int_0^R \Psi_{ln}(r)  dr, 
\label{eq:B_ln}
\end{equation}
where $\Psi_{ln}$ in the {\it neutrino survival eigenfunction}.
$\Psi_{ln}$ reads
\begin{equation} 
\Psi_{ln}=C_\phi^{-1}\; 
\left( \frac{\delta T}{T}\right)_{ln}  \Phi \; \rho\; 4\pi r^2, 
\label{eq:Psi_ln}
\end{equation}
where  $ C_\phi $ is a normalization constant given by  
$ C_\phi= \int \Phi  (r) 4 \pi \rho (r) r^ 2  \;dr $.
Figure~\ref{figu2} presents the function $|\Psi_{ln}|$  for several modes, and 
Figure~\ref{figu3} and Table~\ref{tab:1} show the  values  of ${\cal B}_{ln}$. 
$\Psi_{ln}$ results from the 
superposition of the eigenfunction  $\delta T_{ln}$ with $\Phi(r)$, 
as schematically illustrated in Figure~\ref{figu1}. 
The $|\Psi_{ln}|$ functions have an identical shape for all the modes, 
the difference being only in the magnitude of $|\Psi_{ln}|$, 
which depends on the magnitude of $\delta T_{ln}$
within the region where the $^8$B neutrino flux is produced  (see Figure~\ref{figu2}).
The contribution of  $\delta T_{ln}$ is more important in the case of low $n$ global  modes,
for which the amplitude of $\delta T_{ln}$ is large in the Sun's core, which 
leads to a large value of ${\cal B}_{ln}$. 
Furthermore, in general, radial  acoustic modes have a more important impact on the  $^8$B neutrino 
flux than other global modes,  and consequently produce a relatively larger value of
${\cal B}_{ln}$,  as shown in Figure~\ref{figu3} and Table~\ref{tab:1}. 

Clearly, the impact  on the neutrino fluxes is larger for the $f$-modes 
and the $p_n$ modes of low degree. The strong impact on the $f$-modes of low degree 
is related to the fact that 
their temperature eigenfunctions have their maximum near the center of the Sun (see Figure~\ref{figu1}), 
which also corresponds  
to the maximum of the kinetic energy of the mode, as shown by~\citet{2000A&A...353..775P}.
These $f$-modes are quite distinct from $f$-modes of high degree which have their maximum amplitude
located near the surface~\citep{1997ApJ...489L.197S}.
In fact, in the solar core these low degree $f$-modes have an amplitude 
with the same order as the low order gravity modes.

Among the $p_n$ modes, we notice the fact that  
the radial modes with $n$ equal to $1$, $2$ and $3$ have a larger impact on the  $^8$B neutrino flux 
than other $p_n$ modes. This is due to the fact that the temperature eigenfunction of these modes 
has a relatively large value within the region of production of the $^8$B neutrinos (see Figure~\ref{figu1}).

Presently, the excitation of acoustic and gravity waves is attributed 
to the random motions of convective elements in the upper layers,
due to the Reynolds stress tensor or the advection of turbulent fluctuations 
of entropy~\citep[e.g.,][]{1994ApJ...424..466G,2008A&A...489..291S}. 
The prediction of their amplitudes in the solar core is highly 
unreliable due to the uncertainty associated with the excitation and dumping mechanisms. 
Nevertheless, we make a qualitative estimation of their amplitudes using observational information from  
acoustic modes.

The amplitude for neutrino flux fluctuations $|\Delta \phi_{ln}/\phi_o|_{\rm max}$
can be computed from the amplitude of the temperature eigenfunction ${\cal A}_{ln}$. 
From Equation (\ref{eq:delta_phi_t}),      
one obtains $|\Delta \phi_{ln}/\phi_o|_{\rm max}=25.5 \; {\cal A}_{ln} {\cal B}_{ln}$ (with $\zeta_o=25.5$).
Following a procedure used to compute the surface amplitude of oscillations,
${\cal A}_{nl}$ is estimated from ${\cal A}_{nl}\approx 2/3\;\delta \rho/\rho$  
(for $\Gamma_3 \sim 5/3$) and  $\delta \rho/\rho =v_{\rm ex}/c_{\rm s} $, 
where $v_{\rm ex}$ is the fluid velocity related with the excitation of acoustic modes 
and $c_{\rm s}$ is the adiabatic sound speed~\citep{1959flme.book.....L}. 
On the Sun's surface, $v_{\rm ex} \sim 10\; {\rm cm\; s^{-1}}$~\citep[e.g.,][]{2011JPhCS.271a2049G}
and $c_{\rm s}$ as computed from the SSM is $\sim 7\times 10^{5}\;{\rm cm\; s^{-1}}$~\citep[]{2012RAA....12.1107T}. 
It follows that ${\cal A}_{nl} \sim  10^{-5}$ and 
$|\Delta \phi_{ln}/\phi_o|_{\rm max}\sim 2.5 \; 10^{-4}\; {\cal B}_{ln} $.
If we choose  $ {\cal B}_{ln}\sim 1 $ (see Table~\ref{tab:1}) then $|\Delta \phi_{ln}/\phi_o|_{\rm max} $
has an amplitude of the order of 0.01\%.
The largest values of $|\Delta \phi_{ln}/\phi_o|_{\rm max}$ correspond to the 
$f$-modes and $p_n$ ($n=1,2,3$) radial modes and $p_1$ dipole mode.  
We note that this small $|\Delta \phi_{ln}/\phi_o|_{\rm max}$ value could be larger by a few orders of magnitude, as the exact mechanism of excitation is not known and  several other processes 
can affect its estimation, such as the rotation, non-adiabaticity and  strong-magnetic fields~\citep[e.g.,][]{1991ApJ...370..752G}.

%%%%%%%%%%%%%%%%%%%%%%%%%%%%%%%%%%%%%%%%%%%%%%%%%%%%%%%%%%%%%%%%%%%%%%%%%%%%%
\section{Summary and Conclusion}
%%%%%%%%%%%%%%%%%%%%%%%%%%%%%%%%%%%%%%%%%%%%%%%%%%%%%%%%%%%%%%%%%%%%%%%%%%%%%

In this letter, we discuss the possibility 
of detecting global acoustic mode oscillations through the 
spectral analysis of the $^8$B neutrino flux time series. 
The acoustic modes present in the  $^8$B neutrino flux fluctuations  
can be classified into two groups, low 
and high order modes, for which the amplitude of the temperature eigenfunctions 
is significant or minimal, respectively near the center of the star. 
The first group of modes  with frequencies in the range of
$250\;\mu{\rm Hz}$ to $500\;\mu{\rm Hz}$ (or with a period in the range 30 minutes to 70 minutes)
has a much more significant impact on the $^8$B neutrino flux than the second one.
This is the case for $f$ modes for $l\ge 2$,  $p_n$ ($n=1,4$) radial modes  and the $p_1$ dipole mode.
Therefore, these modes should be the most visible modes in the $^8$B neutrino flux time series.
 
One of the first attempts to detect temporal fluctuations on the $^8$B neutrino flux time series
was by \citet{2010ApJ...710..540A} looking for oscillations
in the frequency range of $1$--$144\;{\rm day}^{-1}$, and their preliminary observational results were negative.  
Following the results obtained in this work, we recommend that observers   look for
global oscillations and   focus their research in the time interval 
 between 30 and 70 minutes.
There is the clear possibility that the next generation of  
astrophysical neutrino detectors might be able to detect such acoustic perturbations 
in the flux of the different solar neutrino sources~\citep[e.g.,][]{2011PhRvD..83c2010W}. 
 
Although this work concentrates on the analysis of the $^8$B neutrino flux,
the $^7$Be, $^{15}$O and $^{17}$F neutrino fluxes should have 
very similar flux variations  because these neutrinos 
are emitted in the same nuclear region. Therefore, 
the acoustic modes observed in the $^8$B neutrino flux
should also be observed in the $^7$Be, $^{15}$O and $^{17}$F  neutrino flux time series. 
Preliminary simulations done by~\citet{2007AIPC..944...25C} 
of the SNO+ detector~\citep{2011NuPhS.217...50M} suggest that, after 3 yr of operation,  
the CNO neutrino  rate should be known with an accuracy of 10\%.
Currently, the $^7$Be neutrino flux presents the best hope 
of detecting these acoustic oscillations. 
Monte Carlo simulations performed by~\citet{2011PhRvD..83c2010W}
suggest that there is the potential of the Low Energy Neutrino Astronomy  
experiment  to determine temporal variations with amplitudes  
of the order 0.5\%, covering a period ranging
from tens of minutes to hundred of years.  
 
The discovery of  global acoustic modes of low $n$ 
in solar neutrino fluxes will be of major interest for the solar physics community,  
because  it will allow an increase in the number of observed acoustic modes 
that are sensitive to the inner core of the Sun.
Solar neutrino seismology 
provides a new way to measure the frequency of global acoustic modes of 
low $n$, which have been very difficult to measure using the current helioseismology techniques.

Furthermore, it will also allow the independent  confirmation of the accuracy 
of the frequency measurements of the low  order acoustic modes already 
measured by a few helioseismic instruments (see Table~\ref{tab:1}), such as,
for instance, the $p_1$ radial mode which has been already observed by GOLF~\citep{2000ApJ...537L.143B}.
This is quite interesting as these frequency measurements correspond to eigenfunctions 
that are sensitive uniquely to the core of our star.

If such a discovery is achieved, it will be a significant step toward  obtaining an accurate diagnostic 
of the inner core of the Sun.  In particular, it will  dramatically improve   
the inversion of the speed of sound and density profiles in the 
deepest layers of the Sun's nuclear region.

% % % % % % % % % % % % % % % % % % % % % % % % % % % % % % % % % % % % % % % % % %
\begin{acknowledgments}
This work was supported by grants from "Funda\c c\~ao para a Ci\^encia e Tecnologia" 
and "Funda\c c\~ao Calouste Gulbenkian".
The author thanks Sylvaine Turck-Chi\`eze for   fruitful discussions 
which have improved the contents and clarity of the paper, as well as 
the authors of CESAM and ADIPLS  for making their codes publicly available.
\end{acknowledgments} 
%\newpage
\bibliography{abclib}
%\bibliography{mnillib}
%\bibliography{mnlibneut}
% % % % % % % % % % % % % % % % % % % % % % % % % % % % % % % % % % % % % % % % % % % % % % % % %
% % % % % % % % % % % % % % % % % % % % % % % % % % % % % % % % % % % % % % % % % %

\end{document}